\documentstyle[12pt]{article}
\begin{document}
\title {Renormalization of Non-locally Regularized BRST-anti-BRST theories}
\author{V. Calian, G. Stoenescu
\\University of Craiova, Faculty of Physics \\ 
13 A. I. Cuza, Craiova 1100, Romania} 
\date{}
\maketitle

\begin{abstract}
An extension of the non-local regularization scheme is formulated in the
Sp(2) symmetric Lagrangian BRST quantization framework. It generates a
systematic treatment of the anomalous quantum master equations and allows to 
subtract the divergences as well as to calculate genuine higher loop 
BRST and anti BRST anomalies.

\end{abstract}

\newpage
\section{Introduction}

The quantization methods based on the BRST symmetry \cite{1}-\cite{5} are
acknowledged as powerful techniques in studying gauge theories. In addition,
the Sp(2) extended version of the Lagrangian formalism \cite{6}, \cite{10}
provided the appropriate treatment of both BRST and anti-BRST symmetries.
However, the quantization stage, the anomalies and renormalization problems
were less addressed in this second framework until now, although several
renormalization procedures (using BPHZ \cite{9} or dimensional regularization 
\cite{16}) have been completed, but only in the field-antifield case.

In this paper we propose a generalization of the non-local regularization 
scheme \cite{7},\cite{8},\cite{14} to the Sp(2) symmetric Lagrangian 
formulation of the BRST
quantization procedure,including a more complete treatment of the 
renormalization stage and enforcing the $\overline{\Delta }^{a}$ -operators 
to act only on non-local expressions at regularized level. The
main difficulty of such a regularization technique arises from the
degenerate antibracket structure of the formalism and from the non-symmetric
roles of the antifields $\phi _{\alpha a}^{*},\overline{\phi }_{\alpha }$
which correspond to every given field $\phi ^{\alpha }$.

This problem can be solved by two types of strategies. One of them \cite{13}
was previously constructed in detail, based on the degeneration itself, by
keeping the entire set of fields and antifields $\left( \phi ^{\alpha },\phi
_{\alpha a}^{*},\overline{\phi }_{\alpha }\right) $ as required by the
standard method. On the other hand, one can rely on the larger but perfectly
symmetric structure of the phase space \cite{11} endowed with Darboux type
coordinates where $\phi ^{\alpha },\pi _{\alpha a}$ (including thus the
Nakanishi-Lautrup fields) are considered fields and $\phi _{\alpha a}^{*},%
\overline{\phi }_{\alpha }$  are treated as the conjugated antifields.

In these coordinates, the antibrackets and the $\Delta ^{a},V^{a}$ operators
are defined as in \cite{11}. 
The generalization of the proposed regularization scheme involves a set of
major differences with respect to the field-antifield case. The anomalous
master equation structure looses its well known and useful symmetry
exhibited in the BV formalism. The set of generalised Wess-Zumino conditions
is more complex and the restrictions imposed to the counter-terms have more
involved cohomological consequences.

The paper is organized as follows. In the second section, we give the
non-local form of the action which may be used in the perturbation theory
calculations for a gauge theory with a known proper solution of the Sp(2)
classical master equations. In the section 3, the general algorithm for
treating the genuine BRST anti BRST higher order anomalies is provided, 
after the divergences subtraction is completed, at each perturbative stage. 
The paper concludes with the discussion of certain open problems and 
consequences.

\section{Sp(2) Non-local regularization}

Our starting point is a gauge theory having the proper solution 
$S\left(\phi ,\phi _{a}^{*},\overline{\phi}, \pi_{\alpha a}\right) $ of the 
classical master equations ($a=1,2$) in Sp(2) Lagrangian formalism:
\begin{equation}
\frac{1}{2}\left( S,S\right) ^{a}+V^{a}S=0
\end{equation}
which may be written as the sum of a free and interaction
contributions:
\begin{equation}
S\left( \phi ,\phi _{a}^{*},\overline{\phi}, \pi_{\alpha a}\right) =F\left( \phi \right) +I_{cl}\left( \phi
,\phi _{a}^{*},\overline{\phi}, \pi_{\alpha a}\right)   
\label{46}
\end{equation}
where 
\begin{equation}
F\left( \phi \right) =\frac{1}{2}\phi ^{A}F_{AB}\phi ^{B}  \label{47}
\end{equation}
while the ''interaction'' term is supposed to be analytic in the
neighbourhood of $\phi ^{A}=0$. The existence of a local solution of this
type is guaranteed by well known theorems (see \cite{15}. \cite{17}).

The quantum action is then perturbatively developed as: 
\begin{equation}
W=F+I_{cl}+\sum\limits_{p\geq 1}\hbar ^{p}M_{p}=F+I  
\label{48}
\end{equation}
where $I\left( \phi ,\phi _{a}^{*},\overline{\phi},\pi_{a}\right) $ is 
the quantum generalization of
the interaction part, which will include cut-off dependent terms.

We define the smoothing operator $\epsilon $ and the cut-off parameter $%
\Lambda ^2$, generating a second order derivative regulator of the type \cite{7}: 
\begin{equation}  \label{49}
R_B^A=\left( T^{-1}\right) ^{AC}F_{CB}
\end{equation}
where the symmetric operator $\left(T^{-1}\right)^{AB}$ does not depend on the fields, while:
\begin{equation} 
\label{50}
\epsilon _B^A=\exp \left( \frac{R_B^A}{2\Lambda ^2}\right)
\end{equation}

The phase space is temporarily enlarged by adding the so-called ''shadow''
fields and antifields $\{ \varphi ,\varphi _a^{*},\overline{\varphi},
\phi_{\varphi; a}\}$ 
having the same statistics as the original fields the antibracket structures 
being extended as well.

The ''shadow'' propagator $O$  corresponding to these fields is, in perfect
agreement with \cite{9}, 
the auxiliary quantum action being then given by: 
\begin{equation}
\overline{W}\left( \phi ,\phi _a^{*},\overline{\phi},\pi_a,\varphi ,
\varphi _a^{*},\overline{\varphi},\pi_{\varphi a}\right) =
F\left(\left( \epsilon ^{-1}\right) \phi \right) -\frac 12\varphi
^AO_{AB}^{-1}\varphi ^B+I\left( \phi ,\phi _a^{*},\overline{\phi},\pi_a,
\varphi ,\varphi _a^{*},\overline{\varphi},\pi_{\varphi a} \right)
\end{equation}

One can easily check that the  BRST-anti-BRST symmetry that leaves the 
classical action $\overline{S}$ invariant is defined by:
\begin{equation}
\overline{\delta}^{a}\phi^{A}=\left(\epsilon^{2} \right)^{A}_{B}R^{aA}\left(\phi+\varphi \right)
\end{equation} 
\begin{equation}
\overline{\delta}^{a}\varphi^{A}=\left(1-\epsilon^{2} \right)^{A}_{B}R^{aA}\left(\phi+\varphi \right)
\end{equation} 
if the original action is invariant under the transformations:
\begin{equation}
\delta\phi^{A}=R^{A}\left(\phi\right)
\end{equation}

The standard expansion of the type:
\begin{equation}
S=S_{0}+\lambda^{*}_{Aa}R^{Aa}\left({\lambda}\right)+\overline{\lambda}^{A}Y_{A}\left(\lambda\right)+\pi_{\lambda Aa}X^{Aa}+...
\end{equation}
and the relations imposed by the classical master equations on the ``structure
 functions'' \cite{6} allow us to identify new ``coordinates'', obtained from 
the old ones by linear transformations: 
\begin{equation}
\lambda ^{A}=\phi ^{A}+\varphi ^{A}  
\label{52}
\end{equation}
\begin{equation}
\lambda _{Aa}^{*}=\left[ \phi _{Ba}^{*}\left( \epsilon ^{2}\right)
_{A}^{B}+\varphi _{Ba}^{*}\left( 1-\epsilon ^{2}\right) _{A}^{B}\right] 
\label{53}
\end{equation}
\begin{equation}
\overline{\lambda}^{A}=\overline{\phi}^{A}+\overline{\varphi}^{A}
\label{54}
\end{equation}
\begin{equation}
\pi _{\lambda;Aa}^{*}=\left[ \pi _{Ba}^{*}\left( \epsilon ^{2}\right)
_{A}^{B}+\pi _{\varphi;Ba}^{*}\left( 1-\epsilon ^{2}\right) _{A}^{B}\right] 
\label{153}
\end{equation}
Consequently, the action $\overline{W}$ can be written as the sum of two terms:
$W\left( \lambda ,\lambda _a^{*},\overline{\lambda},\pi_{\lambda;a}\right)$ 
and a second one, which depends on  additional coordinates $\ \phi^{A}-\left( \epsilon ^{2}\right) _{B}^{A}\left(\phi ^{B}+\varphi^{B}\right)$, 
quadratically.

The new action is thus obtained by replacing the original fields $\phi ^{A},
\overline{\phi}$
with their smeared versions $\left( \epsilon ^{-1}\right) _{B}^{A}\phi ^{B},
\left( \epsilon ^{-1}\right) _{B}^{A}\overline{\phi} ^{B}$
and by adding the shadow fields contribution defined by the propagator 
$O^{AB}$, to the action. 
The antifields $\phi _{Aa}^{*},\pi_{Aa}$ have to be replaced in the 
interaction functional by 
$\lambda _{Aa}^{*}$ and $\pi_{\lambda;Aa}$, respectively.

The Sp(2) generalization of the results \cite{7} - \cite{9} will therefore 
guarantee that the process does not lead to distortions of the extended 
BRST-anti-BRST structure. The new perturbation theory is equivalent to the 
initial one if the external lines $\varphi$ are excluded. The aim is thus to 
eliminate the closed loops generated by shadow lines. We will accomplish this 
by using the canonical transformations derived in \cite{17} which will set 
$\varphi^{*}, \overline{\varphi}, \pi_{\varphi}$ to zero and keep only 
the fields $\varphi$ equal to their on-shell values.
The solution $\varphi _{q}\left( \phi ,\phi^{*},\overline{\phi},\pi\right)$ 
is then replaced in the auxiliary action 
$\overline{W}\left(\phi ,\phi ^{*},\overline{\phi},\pi,\varphi ,\varphi ^{*},
\overline{\varphi},\pi_{\varphi}\right) $.

The final form of the non-local quantum action is thus the one that has to
be used in the regularized perturbative calculations: 
\begin{equation}
W_{\Lambda }\left( \phi ,\phi ^{*},\overline{\phi},\pi\right) \equiv 
\overline{W}\left( \phi
,\phi ^{*},\overline{\phi},\pi,\varphi _{q}\left(\phi,\phi^{*},
\overline{\phi},\pi\right),\varphi ^{*}=0,
\overline{\varphi}=0,\pi_{\varphi}=0\right)  
 \label{59}
\end{equation}
while the expansion: 
\begin{equation}
W_{\Lambda }=S_{\Lambda }+\sum\limits_{p\geq 1}\hbar ^{p}M_{p,\Lambda }
\label{60}
\end{equation}
gives both the classical action and the counter-terms.

At quantum level, the extended BRST structure and its possible violations 
\cite{13} are described by the regularized version of the
Ward identities: 
\begin{equation}
\frac{1}{2}\left( \Gamma _{\Lambda },\Gamma _{\Lambda }\right)
^{a}+V^{a}\Gamma _{\Lambda }=-i\hbar \left( {\cal A}_{\Lambda}^{a}\cdot
\Gamma _{\Lambda }\right)   \label{61}
\end{equation}
where $\Gamma _{\Lambda }$ is the effective action associated to the
regularized quantum action $W_{\Lambda }$.

The anomalies are still of the form: 
\[{\cal A}_{\Lambda}^a\left( \phi ,\phi ^{*},\overline{\phi},\pi\right) =
\left[ \overline{\Delta }^aW_\Lambda +\frac i{2\hbar }\left( W_\Lambda,
W_\Lambda \right) ^a\right]
\left( \phi ,\phi ^{*},\overline{\phi},\pi\right) = 
\]
\begin{equation}  
\label{62}
={\cal A}_\Lambda ^a\left( \phi +\varphi _q,\phi ^{*}\epsilon ^2,
\overline{\phi},\pi \epsilon^2 \right)
\end{equation}
but this time the action of the $\overline{\Delta }^a$ operators is 
well - defined due to the nonlocality of $W_\Lambda $.

However, both $W_{\Lambda}$ and $\overline{\Delta}^aW_\Lambda$ will contain 
divergent terms. For example, in $\hbar$ - order, one can show that 
$\overline{\Delta}^aS_\Lambda$ will diverge as $\Lambda^{1}$.

Therefore, the regularization can not be removed at this stage, and the limit 
$\lim_{\Lambda\rightarrow \infty }{\cal A }_{\Lambda}^{a}$ is meaningful only 
after the divergences are subtracted and the trivial anomalies are identified.

In what follows, we will not apply a procedure similar to the one which was 
previously used in the field-antifield nonlocal regularization \cite{14}. 

By contrast, the renormalization technique will be manifest in our approach 
and is based on the generating functional for 1PI vertices associated to the
 solution of the quantum master equations. 
 
The ``effective action'' $\Gamma $ and the
complex terms $\left({{\cal A}}^a_{\Lambda}\cdot \Gamma \right) $ which 
incorporate the effects of: local contributions to the anomaly, 
the quantum dressings of the non-trivial anomalies in the previous stages 
and the breakings of the master equation due to the regularization 
non-invariance, are treated, at each perturbative order, as the ones 
generated in the previous step, after the divergences subtraction.

We will therefore define: 
\begin{equation}
\label{10}
\Gamma_{\Lambda} =\sum_{p=0}h^p\Gamma _{\Lambda R_{p-1}}^{\left( p\right) } 
\end{equation}
where each term in the expansion may be explicitly given as: 
\begin{equation}
\label{11}
\Gamma _{\Lambda R_{p-1}}^{\left( p\right) }=\sum_{n=n_p}\Lambda^{-n}\Gamma
_{\Lambda R_{p-1}}^{\left( p\right) n} 
\end{equation}
after $\left( p-1\right) $ steps ($p\geq 1$) have been completed by
eliminating the divergences. The value of the lower limit $n_p$ of the power 
series in $\Lambda^{-1} $ is determined this way . We denoted by: 
$\Gamma _{\Lambda R_0}^{\left( 1\right) }\equiv \Gamma ^{\left( 1\right)
};\Gamma ^{\left( 0\right) n}\equiv S_n;n_1=-1$.

Even the set of equations obtained for higher orders has the formal
aspect of the one given in \cite{12}, it encodes all the contributions 
previously described and involves well-defined expressions: 
\begin{equation}
\label{12}-i\left( \overline{{\cal A}}^a\circ \Gamma _{\Lambda R_{p-1}}^{\left(
p\right) }\right) =\left( \Gamma _{\Lambda R_{p-1}}^{\left( p\right) },\Gamma
_{\Lambda R_{p-1}}^{\left( 0\right) }\right) ^a+\widetilde{V}^a\Gamma
_{\Lambda R_{p-1}}^{\left( p\right) }+\sum_{q=1}^{p-1}\left( \Gamma
_{\Lambda R_{p-1}}^{\left( p\right) },\Gamma _{\Lambda R_{p-1}}^{\left( p-q\right) }\right)^a 
\end{equation}
On the other hand, the inhomogeneous ''consistency conditions'' : 
\begin{equation}
\label{14}\left( \left( \overline{{\cal A}}^{\{a}\circ \Gamma
_{\Lambda R_{p-1}}^{\left( p\right) }\right) ,\Gamma _{\Lambda R_{p-1}}^{
\left( 0\right)
}\right) ^{b\}}+\widetilde{V}^{\{a}\left( \overline{{\cal A}}^{b\}}\circ
\Gamma _{\Lambda R_{p-1}}^{\left( p\right) }\right) =-\sum_{q=1}^{p-1}\left( 
\overline{{\cal A}}^{\{a}\circ \Gamma _{R_{p-1}}^{\left( p\right) },\Gamma
_{R_{p-1}}^{\left( p-q\right) }\right) ^{b\}}
\end{equation}
($p\geq 2$) plays an important role in the divergence  subtraction procedure 
and identification of the genuine anomalies.

Both equations (\ref{12}),(\ref{14}) have to be written in $\Lambda ^{-n}$,
for $n=-n_p,...,-1,0,1,...$ at each value of $p$, while the limit 
$\Lambda^{-1}\rightarrow 0$ may be taken (removing the regularization) only 
when this process does not generate any divergences, i.e. when the terms 
with poles in $\Lambda^{-1}$ have been subtracted.

The starting order of the $\left( \overline{{\cal A}}^a\circ \Gamma _
{\Lambda R_{p-1}}^{\left( p\right)}\right)$ - terms in 
$\Lambda^{-1} $ has to be chosen such that ($p=0$): 

\begin{equation}
\label{15}\left( \Gamma ^{\left( 0\right) },\Gamma ^{\left( 0\right)
}\right) ^a+\widetilde{V}^a\Gamma ^{\left( 0\right) }\equiv \Lambda^{-1}{
\alpha }^{\left( 0\right) a}
\end{equation}
such that, if the regularization is removed in (\ref{15}) , the master 
equation at classical level is recovered.

By denoting $(1/\Lambda)\alpha ^{\left( 1\right) a}$ the $-i\left( \overline{
{\cal A}}^a\circ \Gamma_0^{\left( 1\right) }\right) $ term, the 
equation (\ref{12}) gives:
\begin{equation}
\label{16}
\alpha ^{\left( 1\right) -1a}=\left( \Gamma _{}^{\left( 0\right)0},
\Gamma _{}^{\left( 1\right) 0}\right) ^a+\widetilde{V}^a\Gamma_{}^{
\left( 1\right) 0}+\left( \Gamma _{}^{\left( 1\right) -1},\Gamma_{}^{
\left( 0\right) 1}\right) ^a 
\end{equation}
while the equation (\ref{14}) becomes: 
\begin{equation}
\label{17}
\left( \alpha ^{\left( 1\right) -1\{a}-\left( \Gamma _{}^{\left(
1\right) -1},\Gamma _{}^{\left( 0\right) 1}\right) ^{\{a},\Gamma _{}^{\left(
0\right) 0}\right) ^{b\}}+\widetilde{V}^{\{a}\left( \alpha ^{\left( 1\right)
-1\{a}-\left( \Gamma _{}^{\left( 1\right) -1},\Gamma _{}^{\left( 0\right)
1}\right) \right) ^{b\}}=0 
\end{equation}
and the local contribution to the one loop anomaly is identified: 
\begin{equation}
\label{18}
A_1^a\equiv \alpha ^{\left( 1\right) -1a}-\left( \Gamma
_{}^{\left( 1\right) -1},\Gamma _{}^{\left( 0\right) 1}\right) ^a 
\end{equation}

In order to compute higher orders in $\hbar$ the divergence 
$\Gamma _{}^{\left( 1\right) -1}$ and the trivial anomaly must be
eliminated.

This aim is accomplished by writing the equation (\ref{12}) at order 
$\Lambda^1$: 
\begin{equation}
\label{19}
0=\left( \Gamma _{}^{\left( 0\right) 0},\Gamma _{}^{\left(
1\right) -1}\right) ^a+\widetilde{V}^{\{a}\Gamma _{}^{\left( 1\right) -1} 
\end{equation}
and (\ref{17}) generates: 
\begin{equation}
\label{20}
0=\left( \Gamma _{}^{\left( 0\right) 0},A_1^{\{a}\right) ^{b\}}+
\widetilde{V}^{\{b}A_1^{a\}} 
\end{equation}

The divergence $\Gamma _{}^{\left( 1\right) -1}$ and the trivial anomalies 
may be eliminated by an appropriate $\hbar \Lambda $-dependent BRST- anti- BRST
change of variables as the canonical ones given in \cite{17}, which will
leave a total change in the effective action equal to $- \hbar \Lambda
\Gamma _{}^{\left( 1\right) -1}$ if $\mu _a=\frac{i\Lambda}{2}\epsilon_{ab}s^{
b}\gamma$, 
$\widetilde{\mu }_a=i\Lambda s^{a}\alpha_{1}$ 
where 
$\gamma ,\alpha_1$ 
are given by 
$\Gamma ^{\left( 1\right) -1}=\frac{1}{2}\epsilon_{ab}s^{a}s^{b}\gamma+
\gamma_0$ and 
$A_1^a=s^a\alpha_{1}+\alpha_{01}^a$ 
but with $\hbar ^2/\Lambda ^{-2}$ contributions that will propagate to the 
next level.

The first result on the effective action 
$\Gamma _{\Lambda R_1}$ is then of the following form: 
\begin{equation}
\label{21}
\Gamma _{\Lambda R_1}=\Gamma _{\Lambda R_1}^0+\hbar \sum_{n=0}\Lambda ^{-n}
\Gamma_{\Lambda R_1}^{\left( 1\right) n}+\hbar ^2\sum_{n=-2}\Lambda ^{-n}
\Gamma _{\Lambda R_1}^{\left(2\right) n}+O\left( \hbar ^3\right) 
\end{equation}
while the new $-i\left( \overline{{\cal A}}^a\circ \Gamma _{\Lambda R1}^{\left(
2\right) }\right) $ contain two type of terms: $\Lambda^{-1} \alpha _{R_1}^{\left(
2\right) a}$ and the renormalization dressing of the non-trivial one loop
anomaly $\Lambda^{-1} \overline{\alpha }_{R_1}^{\left( 1\right) a}$. 
It corresponds, at $\Lambda \rightarrow \infty$, to the genuine anomaly 
${\cal A}^{(a)}_1$ which satisfies the homogeneous consistency condition:
\begin{equation}
\left(S,{\cal A}_1^{\{ a}\right)^{b\}}=0
\label{120}
\end{equation}

In the next step of our algorithm, the equation (\ref{12}) is reproduced for 
$\hbar ^2,\Lambda^0,\Lambda ^{1},\Lambda ^{2}$ and we can identify the 
following local contribution to the anomaly: 
\begin{equation}
\label{22}
\begin{array}{c}
\alpha _{R_1}^{\left( 2\right) -1a}+ 
\overline{\alpha }_{R_1}^{\left( 1\right) -1a}=\frac 12\left( \Gamma
_{R_1}^{\left( 1\right) 0},\Gamma _{R_1}^{\left( 1\right) 0}\right) ^a+%
\widetilde{V}^a\Gamma _{R_1}^{\left( 2\right) 0} \\ 
+\left( \Gamma
_{R_1}^{\left( 0\right) 0},\Gamma _{R_1}^{\left( 2\right) 0}\right)
^a+\left( \Gamma _{R_1}^{\left( 0\right) 1},\Gamma _{R_1}^{\left( 2\right)
-1}\right) ^a+\left( \Gamma _{R_1}^{\left( 0\right) 2},\Gamma _{R_1}^{\left(
2\right) -2}\right) ^a
\end{array}
\end{equation}
with: 
\begin{equation}
\label{23}
A_2^a\equiv \alpha _{R_1}^{\left( 2\right) -1a}-\left( \Gamma
_{\Lambda R_1}^{\left( 0\right) 1},\Gamma _{\Lambda R_1}^{\left( 2\right) -1}
\right)^a+\left( \Gamma _{\Lambda R_1}^{\left( 0\right) 2},\Gamma _{
\Lambda R_1}^{\left( 2\right)-2}\right) ^a 
\end{equation}
as well as the ''triviality'' conditions: 
\begin{equation}
\label{24}
0=\left( \Gamma _{}^{\left( 0\right) 0},A_2^{\{a}\right) ^{b\}}+%
\widetilde{V}^{\{b}A_2^{a\}} 
\end{equation}
and: 
\begin{equation}
\label{25}0=\left( \Gamma _{}^{\left( 0\right) 0},\Gamma _{\Lambda R_1}^{
\left(2\right) -2}\right) ^a+\widetilde{V}^a\Gamma _{\Lambda R_1}^{
\left( 2\right) -2} 
\end{equation}
while the $\Lambda ^{1}$ divergence satisfies: 
\begin{equation}
\label{26}
\left( \Gamma _{}^{\left( 0\right) 0},\Gamma _{\Lambda R_1}^{\left(
2\right) -1}\right) ^a+\widetilde{V}^a\Gamma _{\Lambda R_1}^{\left( 2\right)
-1}=\alpha _{R_1}^{\left( 2\right) -2a}-\left( \Gamma _{}^{\left( 0\right)
1},\Gamma _{\Lambda R_1}^{\left( 2\right) -2}\right) ^a 
\end{equation}

Once again, the divergences must be eliminated, one by one, by the same kind of
BRST-anti-BRST transformations on $S$, leading us to a finite 
$\Gamma _{\Lambda R_2}$ and to a modified 
$\Lambda^{-1} \alpha _{R_2}^{\left( 2\right) a}$.The procedure may be further
applied to obtain a completely regularized and subtracted effective action 
$\Gamma _{\Lambda R_\infty }$ which respects the anomalous equations. 

An instructive example of the way our method works is to calculate the BRST and
anti-BRST anomalies, in the first order of the perturbation theory, for the 
theory of $W_2$ gravity. Our procedure then will provide, starting with the 
classical action:
\begin{equation}
\label{100}
S_0=\frac{1}{2\pi}\int d^2x\left(\partial\phi \overline{\partial}\phi -
h \left(\partial \phi \right)^2 \right)
\end{equation}
the anomalous contribution given by:
\begin{equation}
\label{101}
{\cal{A}}_a^1=\int d^2x c^a \partial^3 h
\end{equation}
up to a numerical factor, and which respect the condition (\ref{120}) and are
in full agreement with the result of \cite{18}.

The non-local regularization and renormalization of the Sp(2) symmetric BV 
formalism is thus able to determine the non-trivial higher order anomalies, in 
any order of the perturbation theory.

\section{Conclusions}

A systematic treatment was proposed in this paper for completing the
regularization and renormalization stages of the Sp(2) symmetric
quantisation scheme. The extension of the non-local regularization technique
proved to be effective in solving the higher loop anomaly problems for both
BRST and anti BRST sectors. 

The main role in this procedure is played by the perfectly symmetrical
structure of the extended phase space \cite{11}. However, the existence of
Darboux like coordinates is not enforced by a general theorem in the
infinite dimensional case, such that the method is restricted to the
problems which admit such a construction. The solution is provided by
the complex ''triplectic quantisation'' \cite{10} which in turn extends
the phase space and the hierarchy of theory levels.

On the other hand, the double-cohomology analysis of the conditions imposed
to the classical action and to the counter-terms were not extensively studied
yet and one should expect them to provide major clues in the calculation of higher loop anomalies as well.

\section{Aknowledgement}
One of the authors , V.C., would like to thank the HEP-Th group at Niels Bohr 
Institute where this work has been completed.

\end{document}